\documentclass[12pt,a4paper]{article}

\usepackage{amsmath}%
\usepackage{amsfonts}%
\usepackage{amssymb}%
\usepackage{adjustbox}%
\usepackage{graphicx}%
\usepackage{graphics,color}%
\usepackage{natbib}%
\usepackage{subfigure}
\usepackage[outdir=./]{epstopdf}
\usepackage[plain,noend]{algorithm2e}

\newtheorem{definition}{Definition}[section]

\begin{document}

\def\spacingset#1{\renewcommand{\baselinestretch}%
{#1}\small\normalsize} \spacingset{1}

   \title{\bf On a loss-based prior for the number of components in mixture models}
  \author{Clara Grazian\hspace{.2cm}\\
    University of Oxford and Universit\`a degli Studi ``Gabriele d'Annunzio''\\
    and \\
    Cristiano Villa\\
    University of Kent \\
    and \\
    Brunero Liseo \\
    Sapienza Universit\`a di Roma
    }
    \date{}
  \maketitle

%%%%%%%%%%%%%%%%%%%%%%%%%%%%%%%%%%%%%%%%%%%%%%%%%%%%%%
%%         	Abstract			%%%
%%%%%%%%%%%%%%%%%%%%%%%%%%%%%%%%%%%%%%%%%%%%%%%%%%%%%%
\begin{abstract}
We propose a prior distribution for the number of components of a finite mixture model. The novelty is that the prior distribution is obtained by considering the loss one would incur if the true value representing the number of components were not considered. The prior has an elegant and easy to implement structure, which allows to naturally include any prior information one may have as well as to opt for a default solution in cases where this information is not available. The performance of the prior, and comparison with existing alternatives, is studied through the analysis of both real and simulated data.
\end{abstract}

\noindent%
\vfill

\newpage
\spacingset{1.45} % DON'T change the spacing!

\section{Introduction}\label{sc_intro}

This paper takes a novel look at the construction of a prior distribution for the number of components for finite mixture models. % In the Bayesian framework, the number of components of a finite mixture, say $k$, is brought into the model as an object of inference, therefore uncertainty about its true value is represented by a prior probability distribution.
These models represent a flexible and rich way of modeling data, allowing to extend the collection of probability distribution that can be considered and used. Mixture models have been widely developed and researched upon for over a century. To name a few key contributions, we have \cite{Tal85}, \cite{Neal92}, \cite{MP00}, \cite{mmr05}, \cite{Fur06} and the forthcoming \cite{Celeuxetal18}. Besides the general literature on mixture models, a wide range of applications have been discussed, including genetics and gene expression profiling \citep{Mcetal02,Yeungetal01}, economics and finance \citep{JuaSteel10,Diasetal10}, social sciences \citep{Reynoldsetal00, Handetal07} and more.

The basic idea of a mixture model is to assume that observations $x$ are drawn from a density which is the result of a combination of components
\begin{equation}\label{eq:basicmodel}
x \sim \sum_{j=1}^k \omega_j f_j(\cdot \mid \theta_j),
\end{equation}
where the form of $f_j$ is known for each $j$, while the parameters $\theta_j$ and the weights $\omega_j$ are unknown and have to be estimated. In this work, we assume $k$ to be unknown as well and, in accordance to the Bayesian framework, we assign a prior distribution on it. 

Besides the above approach, which is what we use here, other methods to deal with an unknown $k$ have been presented. One way is based on model selection and consists in fitting mixtures with $k=1,\dots,K$ (for a suitable $K$) and comparing the models through some index, such as the Bayesian information criteria; see, for example, \cite{Baudry10}. Alternatively, one could set a large $k$ and let the weights posterior behaviour to identify which components are meaningful. This is known as an \textit{overfitted mixture model} and the aim is to define a prior distribution which has a conservative property in reducing a posteriori the number of meaningful components \citep{Rousseau11}; \cite{Grazian18} have discussed the same approach by using the Jeffreys prior for the mixture weights conditionally on the other parameters.

Several techniques have been proposed to deal with $k$ through the use of a prior $\pi(k)$: \cite{RichardsonGreen97}, \cite{Stephens00}, \cite{NobileFearn07} and \cite{McCYang08}. A well-known and widely used method is the reversible Markov chain Monte Carlo \citep{Green95} which, due to its non-trivial set up, has led to the search of alternatives. A recent and interesting one is proposed by \cite{MillerHarrison18}, where the model in \eqref{eq:basicmodel} is written as
\begin{flalign}\label{eq:alternativemodel}
& k \sim P(k), \nonumber \\ 
& (\omega_1,\ldots,\omega_k) \sim \mbox{Dir}(\gamma,\ldots,\gamma), \nonumber \\
& Z_1,\ldots,Z_n \sim (\omega_1,\ldots,\omega_k), \nonumber \\
& \theta_1,\ldots,\theta_k \sim H, \nonumber \\
& x_i \sim f_{\theta_{Z_j}}, \nonumber
\end{flalign}
where $P(k)$ is the prior on the number of components defined over the set $\{1,2,\ldots\}$, $H$ is the prior base measure, both the $Z\mbox{s}$ and the $\theta\mbox{s}$ are conditionally independent and identically distributed and the $Z\mbox{s}$ are latent variable describing the component membership. It is then highlighted the parallelism with the stick-breaking representation of the Dirichlet mixture model and how the Dirichlet mixture model samplers can be applied to finite mixtures as well. Both simulations and real data analysis performed in the present work have been obtained by using the Jain-Neal split-merge samplers \citep{JainNeal04,JainNeal07}, as implemented in \cite{MillerHarrison18}.

In terms of the determination of $\pi(k)$, which is the focus of this work, the literature is definitely gaunt. In particular, it appears that there is only one proposed prior for $k$ with a non-informative flavour, that is $k\sim\mbox{Poi}(1)$ \citep{Nobile05}. Although other authors proposed to use a prior proportional to a Poisson distribution, see for example \cite{PhilSmith96} and \cite{Stephens00}, only \cite{Nobile05} gave some theoretical justifications on how to choose the Poisson parameter when there is lack of prior knowledge about $k$. Another option, suitable when there is no sufficient prior information, would be to assign equal prior mass to every value of $k$; however, in the case one would like to consider, at least theoretically, the possibility of having an infinite support for the number of components, this last solution would not be viable or would need a truncation of the support which might influence inference. Finally, the geometric distribution depicts a possible representation of prior uncertainty \citep{MillerHarrison18}, although no discussion is reserved in setting the value of the parameter in a scenario of insufficient prior information for the number of components.

%A particular consideration has to be made about a different approach to estimate the number of components (i.e. clusters). \cite{} propose to use a shrinkage effect on the weights, by appropriately tuning the Dirichlet prior put on them, and consider only the components with an associated weight larger than zero.

%For the remainder of the paper, note the following. 
Although the illustrations we present here will refer to mixtures of univariate and multivariate normal densities, the loss-based prior for $k$ we introduce does not depend on the form of the $f_j$s, therefore suitable for any mixture. Throughout the paper we will adopt, for the weights and the component parameters, the priors proposed in  
\cite{MillerHarrison18}; this will not affect the analysis of the results and the comparisons among different priors for $k$.

\section{Prior for the number of components}\label{sc_theprior}

A finite mixture distribution (or, simply, mixture model) is based on the assumption that a set of observed random variables $x=\{x_1,\ldots,x_n\}$ has been generated by a process that can be represented by a weighted sum of probability distributions. That is
\begin{equation}\label{eq_mixmodel}
g(x \mid k,\omega,\theta) = \sum_{j=1}^k \omega_j f_j(x_i \mid \theta_j), \qquad i=1,\ldots,n
\end{equation}
where $f_j(\cdot|\cdot)$ is the probability distribution of the $j$-th component, $\theta=(\theta_1,\ldots,\theta_k)$, $\theta_j$ is the (possibly vector-valued) parameter of $f_j$ and $\omega=(\omega_1,\ldots,\omega_k)$ are the weights of the components, with $\omega_j>0$ for $\forall j=1, \cdots, k$ and $\sum_{j=1}^k\omega_j=1$. For the model \eqref{eq_mixmodel} the prior can be specified as follows
\begin{equation}\label{eq_totalprior}
\pi(k,\omega,\theta) = P(k)\pi(\omega \mid k)\pi(\theta \mid k).
\end{equation}
As mentioned above, the aim of this paper is to define a prior for $k$, therefore the prior distributions for $\omega$ and $\theta$ will be chosen to be proper ``standard'' priors, minimally informative if necessary; see, for example, \cite{RichardsonGreen97} or \cite{MillerHarrison18}.

The posterior for $k$ is then given by
\begin{equation}\label{eq_kpost}
P(k \mid x) \propto \int f(x \mid k,\omega,\theta)\times\pi(k)\pi(\omega \mid k)\pi(\theta \mid k)\, d\omega\, d\theta.
\end{equation}

It is now fundamental to discuss the support of $k$. Although for practical purposes the range of values $k$ can take is finite, $k=1,2,\ldots,K$, it may be appropriate to define a prior over \color{blue}$\mathbb{N}$. \color{black} In fact, by truncating the support of $k$ there may be possible distortions of the posterior around the boundary, affecting the inferential results. However, it has to be noticed that this is needed when using a uniform prior, since the prior on $k$ must be proper, as proved by \cite{Nobile05}. It seems, therefore, more reasonable to use a proper prior defined on $\mathbb{N}$.

To obtain the loss-based prior for $k$, we employ the approach introduced in \cite{VW15}, where a \emph{worth} is associated to each value of $k$ by considering the loss one would incur if that $k$ were removed, and it is the true number of components. Then, the prior $P(k)$ is obtained by linking the \emph{worth} through the self-information loss function \citep{MF98}. In brief, the self-information loss function associates a loss to a probability statement, say $P(A)$, and it has the form $-\log P(A)$.

To determine the \emph{worth} of a given $k$, we proceed as follows. First, we note that there is no loss in information. In fact, a well-known Bayesian property \citep{Berk66} shows that the posterior distribution of a parameter (if the true parameter value is removed) accumulates, in a Kullback-Leibler sense, on the parameter value that identifies the model most similar to the true one. This is equivalent in minimising the loss in information one would incur. Now, if we consider a mixture with $k$ components, the minimum loss would be measured from any mixture with $k^\prime>k$ components. In fact, the mixture with $k^\prime$ components has more parameters (i.e. uncertainty) than the mixture with $k$ components, meaning that the informational content of the former is larger than the one of the latter and the loss in information is zero. That is, $\mbox{Loss}_{I}(k)=0$. However, should we consider only the loss in information, the resulting prior would be the uniform, that is
\begin{eqnarray*}
-\log P(k) &=& 0 \\
P(k) &\propto& 1.
\end{eqnarray*}

To have a sensible prior, we add a loss component that takes into consideration the increasing complexity of the mixture model. This can be interpreted as the loss in complexity that we incur if we remove the $k$-mixture and it is the true one, and it is related to the number of parameters we avoid in estimating (and the number of extra observations we would need to add, in general, to have reliable estimates). We therefore have $\mbox{Loss}_{C}=k$. As such, the total loss associated to the mixture with $k$ components is given by $\mbox{Loss}(k) = \mbox{Loss}_{I}(k) + \mbox{Loss}_{C}(k) = k$,
yielding
\begin{equation}
\label{eq:prior}
P(k) \propto \exp\left\{-c\cdot k\right\},
\end{equation}
where $c>0$ is included as loss functions are defined up to a constant. Although the prior in \eqref{eq:prior} could be directly used, with the the interpretation that $c$ is a hyper-parameter which allows to control for sparsity, our recommendation is to reparametrise it by setting $p=\exp(-c)$ and assign a suitable prior to $p$. In particular, by having $p\sim\mbox{Beta}(\alpha,\beta)$, the prior for $k$ is a particular beta-negative-binomial, that is a beta-geometric distribution, when the support for $k$ is infinite, as the following Definition \ref{def:infiniteprior} (whose detailed derivation is in the Appendix 1) shows.

\begin{definition}\label{def:infiniteprior}
Consider the prior distribution for the number of components of a finite mixture model, as defined in \eqref{eq:prior}, where we set $p=\exp\{-c\}$ and $k=1,2,\ldots$. If we choose $p\sim\mbox{Beta}(\alpha,\beta)$, with $\alpha,\beta>0$, then
$$P(k|p) = p^{k-1}(1-p),$$
which is a geometric distribution with parameter $1-p$, and
\begin{equation}\label{eq:priorbnb}
P(k) = \frac{\Gamma(\alpha+\beta)}{\Gamma(\alpha)\Gamma(\beta)} \frac{\Gamma(k+\beta-1)\Gamma(\alpha+1)}{\Gamma(k+\alpha+\beta)},
\end{equation}
which is a beta-negative-binomial distribution where the number of failures before the experiment is stopped equal to 1, and shape parameters $\alpha$ and $\beta$.
\end{definition}
The prior in \eqref{eq:priorbnb} is strictly positive on the whole support of $k$. This is a necessary condition \citep{Nobile94} to have consistency on the number of components. In addition, the prior in \eqref{eq:priorbnb} is proper, which is another requirement to yield a proper posterior \citep{Nobile04} when the support is $k=\{1,2,\ldots\}$. On this aspect, as the Jeffreys prior for a geometric distribution is improper, the prior for $k$ will be improper as well. As such, a default choice for $P(k)$ should be chosen on different grounds. In particular, the default choice will not give any preference to particular values of $p$, and this can be achieved by setting $p\sim\mbox{Beta}(1,1)$. The resulting prior is then a beta-negative-binomial with all parameter values equal to one which can be rewritten as
\begin{equation}\label{eq:defaultprior}
P(k) = \frac{1}{k(k+1)}.
\end{equation}

In a more general setting, the parameters $\alpha$ and $\beta$ of the Beta prior on $p$ can be used to reflect available prior information about the true number of components. The expectation of the prior in \eqref{eq:priorbnb} is
\begin{equation}\label{eq:expectation}
\mathop{\mathbb{E}}(k) = \mathop{\mathbb{E}}(\mathop{\mathbb{E}}\{k|p\}) = \mathop{\mathbb{E}}(p^{-1}) = \frac{\alpha+\beta-1}{\alpha-1}, \qquad \mbox{for } \alpha>1,
\end{equation}
while the variance has the form
$$\mathrm{Var}(k) = \mathop{\mathbb{E}}(\mathrm{Var}\{k|p\}) + \mathrm{Var}(\mathop{\mathbb{E}}\{k|p\}) = \frac{\alpha\beta(\alpha+\beta-1)}{(\alpha-2)(\alpha-1)^2}, \qquad \mbox{for } \alpha>2.$$
From equation \eqref{eq:expectation} we see that, as $\beta\rightarrow0$, then $\mathop{\mathbb{E}}(k)\rightarrow1$. So, for a given $\alpha>1$, we have that the hyper-parameter $\beta$ can be interpreted as the quantity controlling how many components in the mixture we want \emph{a priori}. The choice of $\alpha$, among values strictly larger than 2, allows to control the variance of the prior, i.e. how certain (or uncertain) \emph{a priori} we are about the true value of $k$.

If the support for $k$ is finite, say $k=\{1,2,\ldots,K\}$, the prior for the number of components (with $p\sim\mbox{Beta}(\alpha,\beta)$) will have the form:
\begin{equation}\label{eq:priorfinite}
P(k) = \int_0^1 \frac{\Gamma(\alpha+\beta)}{\Gamma(\alpha)\Gamma(\beta)}p^{k+\alpha-2}(1-p)^\beta \frac{1}{1-p^K}\,dp,
\end{equation}
which does not have a closed form. Although the prior in \eqref{eq:priorfinite} can be easily implemented in a Markov Chain Monte Carlo procedure, one has to be careful as its performance might depend on the choice of $K$. Besides this, the prior certainly yields a proper posterior for $k$ and is consistent on the number of components.

\section{Illustrations}
To illustrate the performance of the loss-based prior we have run a simulation study and analysed two data sets. In both cases, we have considered univariate and multivariate scenarios, comparing the proposed prior, under default settings, with current alternatives found in the literature.

Before describing the analysis and illustrate the results, the following clarifications have to be made. First, as the aim of this paper is to propose a novel prior distribution for the number of components, we do not discuss in detail the prior assigned to model weights and to the parameters of the components of the mixture. Second, for the same reason, we limit the examples to mixture of normal densities. In fact, keeping both model and priors relatively straightforward allows to better appreciate any difference in the priors. Finally, the computational algorithm implemented assumes that the maximum number of components in the mixture is 50, so that the uniform prior is defined over $k=\{1,\ldots,50\}$; although the truncation is necessary for the uniform prior only, so to have a proper posterior, the choice of 50 is sufficiently large to not interfere with any of the analysis performed.

\subsection{Real data sets}
In this section we illustrate the performance of the prior by analysing two available data sets. The first is the galaxy data set \citep{Roeder90}, which is considered a benchmark for comparison in the univariate case. We also consider a multivariate case; in particular, the discriminating cancer subtypes using gene expression data set \citep{MillerHarrison18}, which has $n=72$ observations for $d=1081$ variables.

The galaxy data sets contains the velocities of 82 galaxies in the Corona Borealis region. Given that the focus here is on the prior for the number of component, we do not go beyond an already tested set up for the model. In particular, the model used in \cite{RichardsonGreen97} where the components of the mixture are normal densities, i.e. $f_j(x) = N(x|\mu_j,\lambda_j^{-1})$), with independent priors for the parameters, normal densities for the means ($\mu_j\sim N(\mu_0,\sigma_0^2)$) and gamma densities for the precision ($\lambda_j\sim \mbox{Gamma}(a,b)$). We also have $a=2$, $b \sim \mbox{Gamma}(a_0,b_0)$, with $a_0=0.2$, while data-dependent priors are chosen for the remaining hyper-parameters: $\mu_0=(\max{x_i}+\min{x_i})/2$, $\sigma_0=\max{x_i}-\min{x_i}$  and $b_0=10/\sigma_0^2$.

\begin{table}[h]
\centering
\caption{Posterior distribution for $k=1,\ldots,10$ for the galaxy data set using the loss-based prior (LB), the uniform prior (UN) and the Poisson prior with parameter 1 (PO).}
\begin{tabular}{ccccccccccc}
%\hline 
$k$ & 1 & 2 & 3 & 4 & 5 & 6 & 7 & 8 & 9 & 10 \\ 
\hline 
LB & 0 & 0 & 0.18 & 0.24 & 0.22 & 0.16 & 0.10 & 0.05 & 0.03 & 0.01 \\ 
%\hline 
UN & 0 & 0 & 0.06 & 0.13 & 0.19 & 0.20 & 0.16 & 0.11 & 0.07 & 0.04 \\ 
%\hline 
PO & 0 & 0 & 0.58 & 0.31 & 0.09 & 0.01 & 0 & 0 & 0 & 0 \\ 
%\hline
\end{tabular}\label{tab:galaxy}
\end{table}

\begin{figure}[h]
\centering
\includegraphics[scale=0.5]{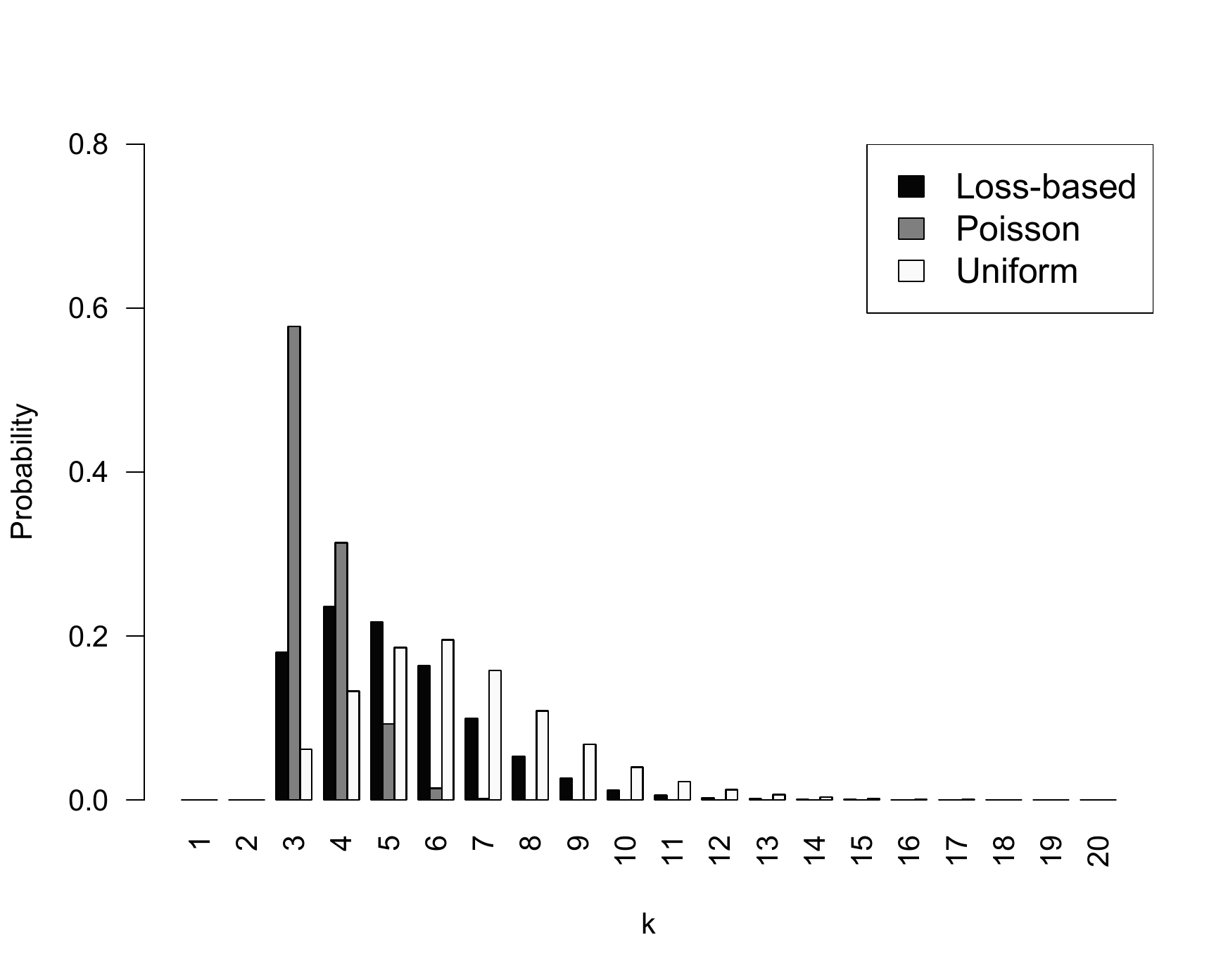}
\caption{Posterior distributions for the Galaxy data set.} 
\label{fig:galaxy}
\end{figure}

Table \ref{tab:galaxy} shows the posterior distribution for the first ten values of $k$ obtained by implementing the loss-based prior, the uniform prior and the Poisson(1) prior. The posterior modes are, respectively, at $k=4$, $k=6$ and $k=3$. The posteriors are also plotted in Figure \ref{fig:galaxy}. There is no unanimous agreement in the number of components in the literature, obviously, and this is supported by the results in Table \ref{tab:galaxy}, which shows estimates of $k$ comparable to what has been identified. However, while the posterior 95\% credible intervals obtained with the loss-based prior and the uniform prior, $[3,9]$ and $[3,12]$ respectively, are sensible, the interval for the Poisson(1) appears to be quite narrow $[3,5]$, excluding values of $k$ previously estimated in the literature. It seems that the loss-based prior provides an intermediate posterior distribution, between the one deriving from the Poisson prior, which is very peaked around $3$, and the one deriving from the uniform prior, which gives non-negligible posterior mass to large values as $12$.

\begin{figure}[h]
\centering
\includegraphics[scale=0.5]{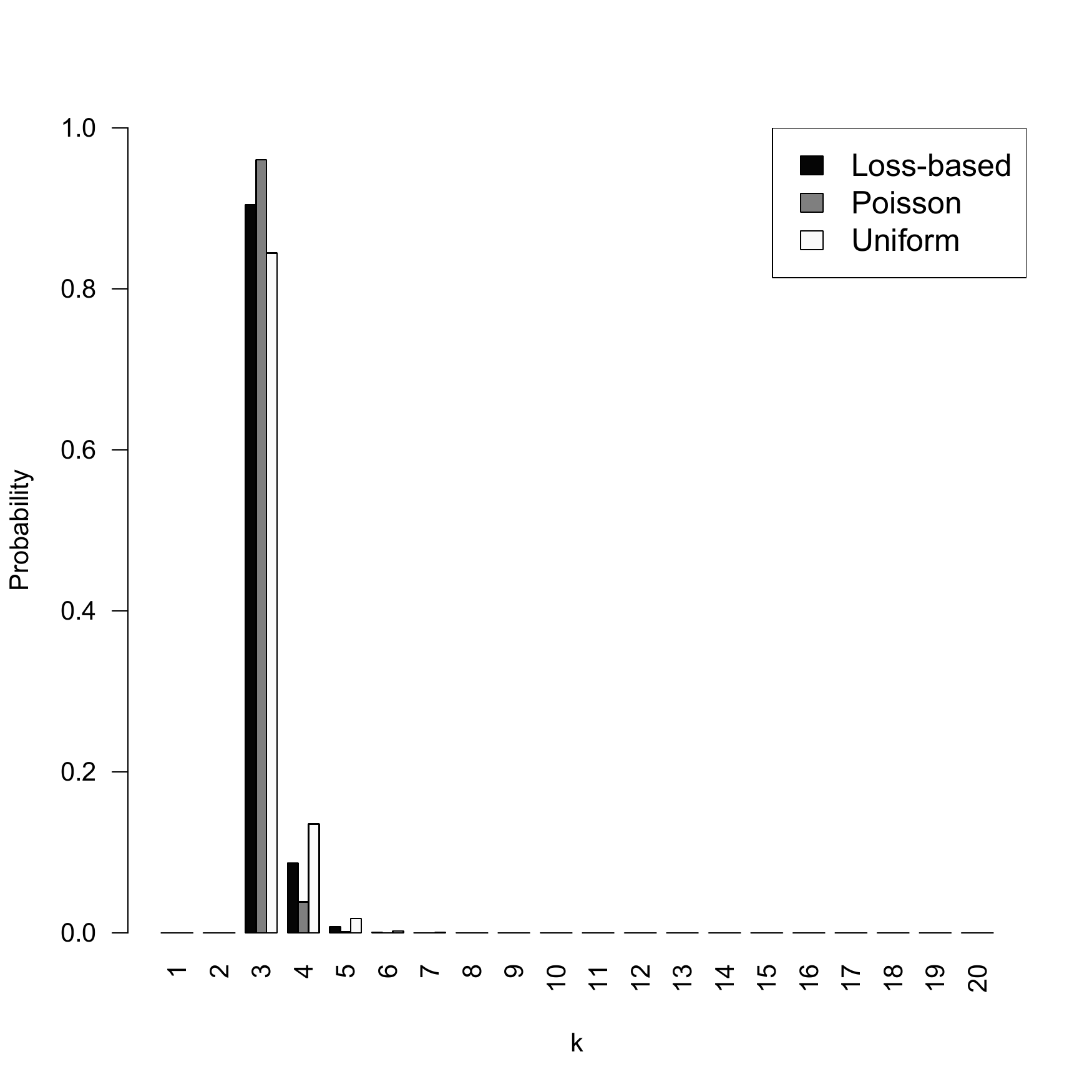}
\caption{Posterior distributions for the cancer gene data set.} 
\label{fig:gene}
\end{figure}

The analysis of the cancer data has given results very similar for the three priors. As Figure \ref{fig:gene} shows, the posterior distributions for $k$ do not differ much, which is supported by the posterior mode, $k=3$, and posterior 95\% credible intervals, $[3,4]$, in all cases. It is obvious that the amount on information about $k$ in the data is sufficiently strong to dominate any of the used priors.

\subsection{Simulation study}\label{sc_simulations}
The simulation study consisted in drawing 100 samples per scenario and compute summaries of indexes of the posterior distributions for the number of components. The details of the simulation study are discussed in the Appendix 2. Briefly, we have considered sample sizes of 50, 100, 500 and 2000. The choice of these sample sizes, for the univariate case, is connected with \cite{McCYang08}, where it has been discussed the challenges of estimating more that $n$ components when the number of observations is actually $n$; thus, a choice of a minimum number of observation larger than the maximum number of observations allows to avoid any pitfall raised in \cite{McCYang08}. For the univariate case, we have considered mixtures of $k=1,2,4,6,12$ components, while for the multivariate case we have considered multivariate mixtures of normal densities of dimension $d=4,8,12$ and $k=3$.
\begin{table}[h]
\centering
\caption{Posterior average mode and average 95\% credible interval (in brackets) for univariate mixtures. The value have been obtained on 100 draws per scenario.}
\begin{tabular}{cclclcl}
	& \multicolumn{6}{c}{Number of components} \\
$n$	& \multicolumn{2}{l}{$k=2$}	&	\multicolumn{2}{l}{$k=4$}	&	\multicolumn{2}{l}{$k=6$}		\\
\hline
50	&	2.07	&	(2.00, 5.28)	&	4.23	&	(2.47, 10.77)	&	1.72	&	(1.00, 6.05)	\\
100	&	2.02	&	(2.00, 4.54)	&	3.26	&	(2.56, 6.38)	&	2.76	&	(1.41, 6.88)	\\
500	&	2.00	&	(2.00, 3.47)	&	3.38	&	(3.07, 4.99)	&	5.65	&	(3.81, 9.85)	\\
2000&	2.03	&	(2.00, 3.15)	&	4.00	&	(3.91, 5.16)	&	7.06	&	(5.74, 11.23)	\\ 
\end{tabular}
\label{tab:LBsummaryUnivariate}
\end{table}
In Table \ref{tab:LBsummaryUnivariate} we have reported the posterior average mode and average 95\% credible intervals for the univariate case. Note that the lower and upper limits of the average posterior credible intervals have been obtained by averaging, respectively, the 100 lower and upper intervals for the 100 samples. We have omitted here the case where $k=1$ and $k=12$, as well as we have considered only one mixture for $k=2$ and $k=3$. Full results are available in the Appendix 2. We note that, as one would expect, the results improve as the sample size $n$ grows, and this is reflected in the values of the posterior (average) mode and on the size of the posterior (average) credible interval. Although for $k=6$ the posterior mode appears not to concentrate on the true number of components, in particular for $n=50$ and $n=100$, the true $k$ is within the credible intervals.

The results concerning mixture of three multivariate normal densities are shown in Table \ref{tab:LBsummaryMultiv}. We again note an improvement on the (average) posterior mode and on the (average) posterior 95\% credible interval as $n$ increases. There is though an improvement on the interval size as the dimensionality increases. It appears that, for a given sample size, the prior is more accurate in detecting the right number of components when $d$ increases, and this is in line with the results obtained for the analysis of the cancer gene data set.
\begin{table}[h]
\centering
\caption{Posterior average mode and average 95\% credible interval (in brackets) for multivariate mixtures of three components ($k=3$). The value have been obtained on 100 draws per scenario.}
\begin{tabular}{cclclcl}
	& \multicolumn{6}{c}{Number of dimensions} \\
$n$	& \multicolumn{2}{l}{$d=4$}	&	\multicolumn{2}{l}{$d=8$}	&	\multicolumn{2}{l}{$d=12$}		\\
\hline
50	&	2.05	&	(2.00, 3.54)	&	2.11	&	(2.00, 3.26)	&	2.06	&	(2.00, 3.13)	\\
100	&	2.61	&	(2.18, 3.87)	&	2.56	&	(2.22, 3.71)	&	2.24	&	(2.11, 3.33)	\\
500	&	3.01	&	(3.00, 3.91)	&	3.00	&	(3.00, 3.26)	&	3.00	&	(3.00, 3.08)	\\
2000&	3.00	&	(3.00, 3.14)	&	3.00	&	(3.00, 3.02)	&	3.00	&	(3.00, 3.00)	\\
\end{tabular}
\label{tab:LBsummaryMultiv}
\end{table}

\section{Discussion}\label{sc_discussion}
To make inference on the number of components in a finite mixture model using the Bayesian framework, one may consider an infinite support. This allows to avoid any potential dependence from the maximum number of components arbitrarily set, giving at the same time more versatility and elegance. The loss-based prior provides a flexible solution to the problem as it allows to incorporate any prior information one may have as well as opting for a default solution in scenario or actual or alleged prior ``ignorance''. In this work, we have shown that, in a setting of limited information, the prior chosen for the number of components influences the posterior distribution; in particular, the uniform prior, which is often used as a default prior, does not seem to be conservative, allowing the posterior distribution to be quite spread towards the possible values. In terms of inference, some level of conservativeness should be preferred, given the fact that the complexity of the inferential problem explodes with the number of meaningful components. On the other hand, the Poisson(1) prior seems to be too conservative, so that, as it is evident in the simulation study, the true value may not even included in the posterior credible interval.

Analysis on both real and simulated data shows that the loss-based prior represents a good compromise between having a prior which excessively penalises for complexity (i.e. the Poisson prior with parameter equal to one) and the uniform prior which suffers from theoretical and implementation weaknesses.

\section*{Appendix 1: Derivation of Definition 1}

Starting from $P(k)\propto\exp\{-c\cdot k\}$, we set $p=e^{-c}$, obtaining
\begin{align}
P(K|p) &= \frac{p^k}{\sum_{m=1}^k p^m} \nonumber \\
&= \frac{p^k}{p/(1-p)} \nonumber \\
&= p^{k-1}(1-p), \tag{S1}\label{eq:geom}
\end{align}
which is a geometric distribution defined over the positive integers and with parameter $(1-p)$. Given the prior $p\sim\mbox{Beta}(\alpha,\beta)$, the marginal for the number of components is given by
\begin{align}
P(k) &= \int_0^1 p^{k-1}(1-p) \pi(p) \,dp \nonumber \\
&= \int_0^1 p^{k-1}(1-p) \frac{1}{\mbox{B}(\alpha,\beta)}p^{\beta-1}(1-p)^{\alpha-1} \,dp \nonumber \\
&= \frac{1}{\mbox{B}(\alpha,\beta)} \int_0^1 p^{k+\beta-2}(1-p)^{\alpha} \,dp \nonumber \\
&= \frac{\Gamma(\alpha+\beta)}{\Gamma(\alpha)\Gamma(\beta)}\frac{\Gamma(k+\beta-1)\Gamma(\alpha+1)}{\Gamma(k+\alpha+\beta)}, \tag{S2}\label{eq:bnb}
\end{align}
where $\mbox{B}(\alpha,\beta)$ is the beta function. Therefore, the marginal distribution for $k$ in \eqref{eq:bnb} is beta-negative-binomial with the parameter expressing the number of failures before stopping equal to 1, and shape parameter $\alpha,\beta>0$. In fact, the generic beta-negative-binomial defined is defined over $k=0,1,\ldots$, and has probability mass function
\begin{align*}
f(k|\alpha,\beta,r) &= \binom{r+k-1}{k} \frac{\Gamma(\alpha+r) \Gamma(\beta+k)\Gamma(\alpha+\beta)}{\Gamma(\alpha+r+\beta+k)\Gamma(\alpha)\Gamma(\beta)} \\
&=\frac{\Gamma(r+k)}{k! \Gamma(r)} \frac{\Gamma(\alpha+r) \Gamma(\beta+k)\Gamma(\alpha+\beta)}{\Gamma(\alpha+r+\beta+k)\Gamma(\alpha)\Gamma(\beta)},
\end{align*}
which, when $r=1$, becomes
\begin{align*}
f(k|\alpha,\beta,1) &=\frac{\Gamma(1+k)}{k! \Gamma(1)} \frac{\Gamma(\alpha+1) \Gamma(\beta+k)\Gamma(\alpha+\beta)}{\Gamma(\alpha+1+\beta+k)\Gamma(\alpha)\Gamma(\beta)} \\
&=\frac{\Gamma(\alpha+\beta)}{\Gamma(\alpha)\Gamma(\beta)} \frac{\Gamma(k+\beta)\Gamma(\alpha+1)}{\Gamma(\alpha+1+\beta+k)} \frac{k!}{k!}.
\end{align*}
Given that we have $k$ defined over the set of strictly positive integer, we rewrite in terms of $k=(k-1)$, obtaining
\begin{equation*}
f(k|\alpha,\beta,1) =\frac{\Gamma(\alpha+\beta)}{\Gamma(\alpha)\Gamma(\beta)} \frac{\Gamma(k+\beta-1)\Gamma(\alpha+1)}{\Gamma(\alpha+\beta+k)},
\end{equation*}
which has the same form as \eqref{eq:bnb}, retrieving the result. We note that the distribution in \eqref{eq:bnb} is a particular case of the beta-negative-binomial, called the beta-geometric distribution.
%------------------------------------------------------------------------------------------------

\section*{Appendix 2: Simulation study}
The simulation study has been performed by drawing 100 samples from different mixture models, encompassing univariate and multivariate mixtures. We have considered the following sample sizes: $n=50,100,500,200$ and, for the multivariate cases, we have considered distributions of dimension $d=4,8,12$.

As the focus of the paper is about making inference on the number of components in a finite mixture, we have not given particular emphasis on the prior distributions for the mixture weights and parameters, as well as the computational techniques. In the univariate case, the models considered \citep{RichardsonGreen97} have normal components, $N(x|\mu_j,\lambda_j^{-1})$, with the following independent priors on the parameters:
\begin{eqnarray*}
\mu_j &\sim& N(\mu_0,\sigma_0^2)\\
\lambda_j &\sim& \mbox{Gamma}(a,b).
\end{eqnarray*}
The hyper-parameters of the normal prior have been set as follows:
$$\mu_0 = (\max x_i + \min x_i)/2 \qquad \mbox{and} \qquad \sigma_0=\max x_i - \min x_i,$$
while for the Gamma prior we have set:
$$a=2 \qquad \mbox{and} \qquad b\sim\mbox{Gamma}(0.2,10/\sigma_0^2).$$
For the multivariate case, for each component $t=1,\ldots,d$, we have $N(\mu_t,\lambda_t^{-1})$, with
\begin{eqnarray*}
\lambda_t &\sim& \mbox{Gamma}(d,h) \\
\mu_t|\lambda_t &\sim& N(0,(w\lambda_i)^{-1}).
\end{eqnarray*}
We have set $d=h=w=1$.
To perform the analysis, given the non-triviality of the Markov Chain Reversible Jump, we have opted for the algorithm implemented in \cite{MillerHarrison18}, included the code provided by the authors.

Finally, each generated sample has been analysed by considering the proposed loss-based prior as well as two options available in the literature (which have a non-informative flavour): the uniform prior and the Poisson(1) \citep{Nobile05}. The algorithm implemented considers the maximum number of components equal to 50, so that the uniform prior for $k$ is defined over $k=\{1,\ldots,50\}$. Note that the truncation is necessary when the uniform prior is employed to ensure a proper posterior, while id does not impact the performance of the other two priors. In any case, the truncation point is sufficiently large to not impact the analysis of both real and simulated data.

For the univariate case, we have considered the mixtures as in Table \ref{tab:listuniv}, while for the multivariate, we have considered the three-component mixtures of the following form:
$$x_i\sim \frac{1}{3}N(m,I)+\frac{1}{3}N(0,I)+\frac{1}{3}N(-m,I),$$
with $m=(3/\sqrt{d},\ldots,3/\sqrt{d})$, for $d=4,8,12$. This choice \citep{MillerHarrison18} will prevent the posterior from concentrating too quickly.
\begin{table}[h]
\caption{List of univariate mixture models.}\label{tab:listuniv}
\centering
\begin{tabular}{ccc}
%\hline 
$k$ & Name & Mixture model \\ 
\hline 
1 & $M_1$ & $x_i\sim N(0,1)$ \\ 
%\hline 
2 & $M_{2a}$ & $x_i\sim \frac{1}{2}N(0,1)) + \frac{1}{2}N(6,1)$ \\ 
%\hline 
2 & $M_{2b}$ & $x_i\sim \frac{2}{3} N(0,1)) + \frac{1}{3} N(6,1)$ \\ 
%\hline 
2 & $M_{2c}$ & $x_i\sim \frac{1}{2} N(0,1)) + \frac{1}{2} N(0,0.15)$ \\ 
%\hline 
4 & $M_{4a}$ & $x_i\sim \frac{1}{4} N(1.5,1)) + \frac{1}{4} N(-1.5,1) + \frac{1}{4} N(4.5,1) + \frac{1}{4} N(-4.5,1)$ \\ 
%\hline 
4 & $M_{4b}$ & $x_i\sim \frac{1}{8} N(-3,1)) + \frac{1}{8} N(3,1) + \frac{1}{2} N(0,0.1) + \frac{1}{4} N(6,2)$ \\ 
%\hline 
6 & $M_6$ & $x_i\sim \sum_{j=0}^5 \frac{1}{6} N(3j,1)$ \\ 
%\hline 
12 & $M_{12}$ & $x_i\sim \sum_{j=0}^{11} \frac{1}{12} N(3j,1)$ \\ 
%\hline 
\end{tabular}
\end{table} 

To remove any possible effect of the priors on the mixture weights and parameters, we have applied a large burn-in period. That is, we have run the simulation for 100,000 iteration and kept the last 1,000 only. As such, convergence has been ensured and the comparison between the prior for the number of components is reliable and meaningful. Moreover, for computational purposes, we have considered $k=1,2,\ldots,50$.

For each scenario, we have computed the average (over the 100 draws) of the posterior mode, as well as the average 95\% credible interval; the lower and upper limits of these intervals have been obtained by averaging, respectively, the 100 lower limits and the 100 upper limits of the posterior credible intervals. The results for the univariate cases are reported from Table \ref{tab:univK1} to Table \ref{tab:univK12}, while for the multivariate, from Table \ref{tab:univk3d4} to Table \ref{tab:univk3d12}.

\begin{table}[h]
\centering
\caption{Posterior indexes for scenario $M_1$.}\label{tab:univK1}
\begin{tabular}{ccccccc}
%\hline 
 & \multicolumn{2}{c}{Loss-Based} & \multicolumn{2}{c}{Uniform} & \multicolumn{2}{c}{Poi(1)} \\ 
\hline 
$n$ & Mode  & C.I. & Mode  & C.I. & Mode  & C.I. \\ 
\hline 
50 & 1.02  & (1, 3.18) & 1.10  & (1, 6.43) & 1.05  & (1, 2.62) \\ 
%\hline 
100 & 1.01  & (1, 2.44) & 1.05  & (1, 4.53) & 1.02  & (1, 2.24) \\ 
%\hline 
500 & 1.01  & (1, 2.01) & 1.03  & (1, 2.55) & 1.01  & (1, 2.03) \\ 
%\hline 
2000 & 1.00  & (1, 1.44) & 1.01 & (1, 2.11) & 1.00  & (1, 1.58) \\ 
%\hline 
\end{tabular} 
\end{table}

%\begin{table}[ht]
%\centering
%\scriptsize
%\begin{tabular}{rrrrrrrrrrrrr}
%  \hline
% n & Mn OBJ & Mn UNI & Mn POI & Sd OBJ & Sd UNI & Sd POI & Fr OBJ & Fr UNI & Fr POI & Md OBJ & Md UNI & Md POI \\ 
%  \hline
% 50 & 1.02 & 1.10 & 1.05 & 0.14 & 0.33 & 0.22 & 0.98 & 0.91 & 0.95 & 1.00 & 1.00 & 1.00 \\ 
% 100 & 1.01 & 1.05 & 1.02 & 0.10 & 0.22 & 0.14 & 0.99 & 0.95 & 0.98 & 1.00 & 1.00 & 1.00 \\ 
% 500 & 1.01 & 1.03 & 1.01 & 0.10 & 0.22 & 0.10 & 0.99 & 0.98 & 0.99 & 1.00 & 1.00 & 1.00 \\ 
% 2000 & 1.00 & 1.01 & 1.00 & 0.00 & 0.10 & 0.00 & 1.00 & 0.99 & 1.00 & 1.00 & 1.00 & 1.00 \\ 
%   \hline
%\end{tabular}
%\caption{Univariate: one component - i.e. $(x_i\sim N(0,1))$.}
%\label{tab:univK1}
%\end{table}

\begin{table}[h]
\centering
\caption{Posterior indexes for scenario $M_{2a}$.}\label{tab:univK21}
\begin{tabular}{ccccccc}
%\hline 
 & \multicolumn{2}{c}{Loss-Based} & \multicolumn{2}{c}{Uniform} & \multicolumn{2}{c}{Poi(1)} \\ 
\hline 
$n$ & Mode  & C.I. & Mode  & C.I. & Mode  & C.I. \\ 
\hline 
50 & 2.03  & (2, 5.28) & 2.15  & (2.01, 8.21) & 2.02  & (2, 3.59) \\ 
%\hline 
100 & 2.01   & (2, 4.54) & 2.10  & (2.00, 6.47) & 2.00 & (2, 3.27) \\ 
%\hline 
500 & 2.00  & (2, 3.47) & 2.01 &  (2.00, 4.38) & 2.00  & (2, 3.03) \\ 
%\hline 
2000 & 2.01  & (2, 3.15) & 2.03  & (2.00, 3.56) & 2.01  & (2, 2.85) \\ 
%\hline 
\end{tabular} 
\end{table}

%\begin{table}[ht]
%\centering
%\scriptsize
%\begin{tabular}{rrrrrrrrrrrrr}
%  \hline
% n & Mn OBJ & Mn UNI & Mn POI & Sd OBJ & Sd UNI & Sd POI & Fr OBJ & Fr UNI & Fr POI & Md OBJ & Md UNI & Md POI \\ 
%  \hline
%50 & 2.03 & 2.15 & 2.02 & 0.22 & 0.44 & 0.20 & 0.98 & 0.87 & 0.99 & 2.00 & 2.00 & 2.00 \\ 
%100 & 2.01 & 2.10 & 2.00 & 0.10 & 0.30 & 0.00 & 0.99 & 0.90 & 1.00 & 2.00 & 2.00 & 2.00 \\ 
%500 & 2.00 & 2.01 & 2.00 & 0.00 & 0.10 & 0.00 & 1.00 & 0.99 & 1.00 & 2.00 & 2.00 & 2.00 \\ 
%2000 & 2.01 & 2.03 & 2.01 & 0.10 & 0.17 & 0.10 & 0.99 & 0.97 & 0.99 & 2.00 & 2.00 & 2.00 \\ 
%   \hline
%\end{tabular}
%\caption{Univariate: two components - i.e. $\Big(x_i\sim \frac{1}{2}N(0,1)) + \frac{1}{2}N(6,1)\Big)$.}
%\label{tab:univK21}
%\end{table}

\begin{table}[h]
\centering
\caption{Posterior indexes for scenario $M_{2b}$.}\label{tab:univK22}
\begin{tabular}{ccccccc}
%\hline 
 & \multicolumn{2}{c}{Loss-Based} & \multicolumn{2}{c}{Uniform} & \multicolumn{2}{c}{Poi(1)} \\ 
\hline 
$n$ & Mode  & C.I. & Mode  & C.I. & Mode  & C.I. \\ 
\hline 
50 & 1.47 & (1.12, 4.83) & 1.77  & (1.29, 8.78) & 1.57  & (1.14, 3.38) \\ 
%\hline 
100 & 2.03 & (2.00, 4.41) & 2.09  & (2.00, 6.30) & 2.02  & (2.00, 3.18) \\ 
%\hline 
500 & 2.01  & (2.00, 3.35) & 2.01  & (2.00, 4.28) & 2.00  & (2.00, 3.02) \\ 
%\hline 
2000 & 2.01  & (2.00, 3.16) & 2.06  & (2.01, 3.56) & 2.01  & (2.00, 2.88) \\ 
%\hline 
\end{tabular} 
\end{table}

%\begin{table}[ht]
%\centering
%\scriptsize
%\begin{tabular}{rrrrrrrrrrrrr}
%  \hline
%n & Mn OBJ & Mn UNI & Mn POI & Sd OBJ & Sd UNI & Sd POI & Fr OBJ & Fr UNI & Fr POI & Md OBJ & Md UNI & Md POI \\ 
%  \hline
%50 & 1.47 & 1.77 & 1.57 & 0.50 & 0.47 & 0.50 & 0.47 & 0.73 & 0.57 & 1.00 & 2.00 & 2.00 \\ 
%100 & 2.03 & 2.09 & 2.02 & 0.17 & 0.32 & 0.14 & 0.97 & 0.92 & 0.98 & 2.00 & 2.00 & 2.00 \\ 
%500 & 2.01 & 2.01 & 2.00 & 0.10 & 0.10 & 0.00 & 0.99 & 0.99 & 1.00 & 2.00 & 2.00 & 2.00 \\ 
%2000 & 2.01 & 2.06 & 2.01 & 0.10 & 0.24 & 0.10 & 0.99 & 0.94 & 0.99 & 2.00 & 2.00 & 2.00 \\ 
%   \hline
%\end{tabular}
%\caption{Univariate: two components - i.e. $\Big(x_i\sim \frac{2}{3} N(0,1)) + \frac{1}{3} N(6,1)\Big)$.}
%\label{tab:univK22}
%\end{table}

\begin{table}[h]
\centering
\caption{Posterior indexes for scenario $M_{2c}$.}\label{tab:univK23}
\begin{tabular}{ccccccc}
%\hline 
 & \multicolumn{2}{c}{Loss-Based} & \multicolumn{2}{c}{Uniform} & \multicolumn{2}{c}{Poi(1)} \\ 
\hline 
$n$ & Mode  & C.I. & Mode  & C.I. & Mode  & C.I. \\ 
\hline 
50 & 2.12  & (1.55, 6.53) & 2.79  & (1.83, 10.19) & 2.02  & (1.54, 3.95) \\ 
%\hline 
100 & 2.23  & (2.00, 5.38) & 2.37  & (2.02, 7.64) & 2.09  & (2.00, 3.59) \\ 
%\hline 
500 & 2.01  & (2.00, 3.18) & 2.05  & (2.00, 3.73) & 2.00  & (2.00, 3.02) \\ 
%\hline 
2000 & 2.00  & (2.00, 2.70) & 2.00  & (2.00, 3.03) & 2.00  & (2.00, 2.49) \\ 
%\hline 
\end{tabular} 
\end{table}

%\begin{table}[ht]
%\centering
%\scriptsize
%\begin{tabular}{rrrrrrrrrrrrr}
%  \hline
%n & Mn OBJ & Mn UNI & Mn POI & Sd OBJ & Sd UNI & Sd POI & Fr OBJ & Fr UNI & Fr POI & Md OBJ & Md UNI & Md POI \\ 
%  \hline
%50 & 2.12 & 2.79 & 2.02 & 0.69 & 0.91 & 0.51 & 0.55 & 0.37 & 0.74 & 2.00 & 3.00 & 2.00 \\ 
%100 & 2.23 & 2.37 & 2.09 & 0.51 & 0.65 & 0.29 & 0.80 & 0.70 & 0.91 & 2.00 & 2.00 & 2.00 \\ 
%500 & 2.01 & 2.05 & 2.00 & 0.10 & 0.26 & 0.00 & 0.99 & 0.96 & 1.00 & 2.00 & 2.00 & 2.00 \\ 
%2000 & 2.00 & 2.00 & 2.00 & 0.00 & 0.00 & 0.00 & 1.00 & 1.00 & 1.00 & 2.00 & 2.00 & 2.00 \\ 
%   \hline
%\end{tabular}
%\caption{Univariate: two components - i.e. $\Big(x_i\sim \frac{1}{2} N(0,1)) + \frac{1}{2} N(0,0.15)\Big)$.}
%\label{tab:univK23}
%\end{table}

\begin{table}[h]
\centering
\caption{Posterior indexes for scenario $M_{4a}$.}\label{tab:univK41}
\begin{tabular}{ccccccc}
%\hline 
 & \multicolumn{2}{c}{Loss-Based} & \multicolumn{2}{c}{Uniform} & \multicolumn{2}{c}{Poi(1)} \\ 
\hline 
$n$ & Mode  & C.I. & Mode  & C.I. & Mode & C.I. \\ 
\hline 
50 & 1.57  & (1.03, 6.20) & 2.50  & (1.29, 11.41) & 1.71  & (1.02, 3.74) \\ 
%\hline 
100 & 2.35   & (1.55, 6.60) & 3.04  & (1.96, 9.61) & 2.24  & (1.50, 4.09) \\ 
%\hline 
500 & 3.88  & (3.23, 7.73) & 4.41  & (3.44, 9.30) & 3.70  & (3.13, 5.10) \\ 
%\hline 
2000 & 4.05  & (4.00, 7.46) & 4.35  & (4.00, 8.61) & 4.00 & (4.00, 5.18) \\ 
%\hline 
\end{tabular} 
\end{table}

%\begin{table}[ht]
%\centering
%\scriptsize
%\begin{tabular}{rrrrrrrrrrrrr}
%  \hline
%n & Mn OBJ & Mn UNI & Mn POI & Sd OBJ & Sd UNI & Sd POI & Fr OBJ & Fr UNI & Fr POI & Md OBJ & Md UNI & Md POI \\ 
%  \hline
%50 & 1.57 & 2.50 & 1.71 & 0.62 & 0.88 & 0.50 & 0.01 & 0.13 & 0.00 & 2.00 & 2.00 & 2.00 \\ 
%100 & 2.35 & 3.04 & 2.24 & 0.72 & 0.92 & 0.49 & 0.06 & 0.20 & 0.02 & 2.00 & 3.00 & 2.00 \\ 
%500 & 3.88 & 4.41 & 3.70 & 0.36 & 0.51 & 0.46 & 0.86 & 0.57 & 0.70 & 4.00 & 4.00 & 4.00 \\ 
%2000 & 4.05 & 4.35 & 4.00 & 0.22 & 0.48 & 0.00 & 0.95 & 0.65 & 1.00 & 4.00 & 4.00 & 4.00 \\ 
%   \hline
%\end{tabular}
%\caption{Univariate: four components - i.e. $\Big(x_i\sim \frac{1}{4} N(1.5,1)) + \frac{1}{4} N(-1.5,1) + \frac{1}{4} N(4.5,1) + \frac{1}{4} N(-4.5,1)\Big)$.}
%\label{tab:univK41}
%\end{table}

\begin{table}[h]
\centering
\caption{Posterior indexes for scenario $M_{4b}$.}\label{tab:univK42}
\begin{tabular}{ccccccc}
%\hline 
 & \multicolumn{2}{c}{Loss-Based} & \multicolumn{2}{c}{Uniform} & \multicolumn{2}{c}{Poi(1)} \\ 
\hline 
$n$ & Mode  & C.I. & Mode  & C.I. & Mode  & C.I. \\ 
\hline 
50 & 3.42  & (2.47, 10.77) & 4.84  & (2.96, 17.16) & 2.73  & (2.25, 4.55) \\ 
%\hline 
100 & 3.07   & (2.56, 6.38) & 3.36  & (2.68, 8.46) & 2.88  & (2.40, 4.25) \\ 
%\hline 
500 & 3.34  & (3.07, 4.99) & 3.42  & (3.09, 5.49) & 3.25  & (3.03, 4.28) \\ 
%\hline 
2000 & 4.00  & (3.91, 5.16) & 4.00  & (3.93, 5.28) & 3.97 & (3.89, 4.56) \\ 
%\hline 
\end{tabular} 
\end{table}

%\begin{table}[ht]
%\centering
%\scriptsize
%\begin{tabular}{rrrrrrrrrrrrr}
%  \hline
%n & Mn OBJ & Mn UNI & Mn POI & Sd OBJ & Sd UNI & Sd POI & Fr OBJ & Fr UNI & Fr POI & Md OBJ & Md UNI & Md POI \\ 
%  \hline
%50 & 3.42 & 4.84 & 2.73 & 1.68 & 2.89 & 0.72 & 0.17 & 0.12 & 0.10 & 3.00 & 4.00 & 3.00 \\ 
%100 & 3.07 & 3.36 & 2.88 & 0.88 & 1.09 & 0.69 & 0.18 & 0.23 & 0.09 & 3.00 & 3.00 & 3.00 \\ 
%500 & 3.34 & 3.42 & 3.25 & 0.48 & 0.50 & 0.44 & 0.34 & 0.42 & 0.25 & 3.00 & 3.00 & 3.00 \\ 
%2000 & 4.00 & 4.00 & 3.97 & 0.14 & 0.14 & 0.17 & 0.98 & 0.98 & 0.97 & 4.00 & 4.00 & 4.00 \\ 
%   \hline
%\end{tabular}
%\caption{Univariate: four components - i.e. $\Big(x_i\sim \frac{1}{8} N(-3,1)) + \frac{1}{8} N(3,1) + \frac{1}{2} N(0,0.1) + \frac{1}{4} N(6,2)\Big)$.}
%\label{tab:univK42}
%\end{table}

\begin{table}[h]
\centering
\caption{Posterior indexes for scenario $M_{6}$.}\label{tab:univK61}
\begin{tabular}{ccccccc}
%\hline 
 & \multicolumn{2}{c}{Loss-Based} & \multicolumn{2}{c}{Uniform} & \multicolumn{2}{c}{Poi(1)} \\ 
\hline 
$n$ & Mode & C.I. & Mode  & C.I. & Mode & C.I. \\ 
\hline 
50 & 1.40  & (1.00, 6.05) & 2.36  & (1.06, 11.87) & 1.59  & (1.00, 3.69) \\ 
%\hline 
100 & 2.26   & (1.41, 6.88) & 3.07  & (1.97, 10.21) & 2.09  & (1.44, 4.06) \\ 
%\hline 
500 & 5.18  & (3.81, 9.85) & 5.90  & (4.06, 11.69) & 4.12  & (3.30, 5.96) \\ 
%\hline 
2000 & 6.71  & (5.74, 11.23) & 7.20  & (5.88, 12.33) & 5.84  & (5.50, 7.18) \\ 
%\hline 
\end{tabular} 
\end{table}

%\begin{table}[ht]
%\centering
%\scriptsize
%\begin{tabular}{rrrrrrrrrrrrr}
%  \hline
%n & Mn OBJ & Mn UNI & Mn POI & Sd OBJ & Sd UNI & Sd POI & Fr OBJ & Fr UNI & Fr POI & Md OBJ & Md UNI & Md POI \\ 
%  \hline
%50 & 1.40 & 2.36 & 1.59 & 0.51 & 0.88 & 0.51 & 0.00 & 0.00 & 0.00 & 1.00 & 2.00 & 2.00 \\ 
%100 & 2.26 & 3.07 & 2.09 & 0.48 & 0.86 & 0.29 & 0.00 & 0.03 & 0.00 & 2.00 & 3.00 & 2.00 \\ 
%500 & 5.18 & 5.90 & 4.12 & 0.90 & 0.90 & 0.70 & 0.27 & 0.47 & 0.04 & 5.00 & 6.00 & 4.00 \\ 
%2000 & 6.71 & 7.20 & 5.84 & 0.52 & 0.49 & 0.37 & 0.26 & 0.03 & 0.84 & 7.00 & 7.00 & 6.00 \\ 
%   \hline
%\end{tabular}
%\caption{Univariate: six components - i.e. $\Big(x_i\sim \sum_{j=0}^5 \frac{1}{6} N(3j,1) \Big)$.}
%\label{tab:univK61}
%\end{table}

\begin{table}[h]
\centering
\caption{Posterior indexes for scenario $M_{12}$.}\label{tab:univK12}
\begin{tabular}{ccccccc}
%\hline 
 & \multicolumn{2}{c}{Loss-Based} & \multicolumn{2}{c}{Uniform} & \multicolumn{2}{c}{Poi(1)} \\ 
\hline 
$n$ & Mode & C.I. & Mode & C.I. & Mode  & C.I. \\ 
\hline 
50 & 1.74 & (1.22, 6.46) & 2.49  & (1.43, 11.93) & 1.81  & (1.19, 3.84) \\ 
%\hline 
100 & 2.48   & (1.73, 7.44) & 3.36 & (2.00, 10.91) & 2.30  & (1.63, 4.23) \\ 
%\hline 
500 & 5.12  & (1.73, 7.44) & 5.12  & (4.06, 12.21) & 4.09  & (3.39, 6.01) \\ 
%\hline 
2000 & 10.60  & (7.64, 17.75) & 11.70  & (8.34, 19.69) & 6.81  & (5.81, 8.59) \\ 
%\hline 
\end{tabular} 
\end{table}

%\begin{table}[ht]
%\centering
%\scriptsize
%\begin{tabular}{rrrrrrrrrrrrr}
%  \hline
%n & Mn OBJ & Mn UNI & Mn POI & Sd OBJ & Sd UNI & Sd POI & Fr OBJ & Fr UNI & Fr POI & Md OBJ & Md UNI & Md POI \\ 
%  \hline
%50 & 1.74 & 2.49 & 1.81 & 0.68 & 1.00 & 0.54 & 0.00 & 0.00 & 0.00 & 2.00 & 2.00 & 2.00 \\ 
%100 & 2.48 & 3.36 & 2.30 & 0.76 & 1.03 & 0.58 & 0.00 & 0.00 & 0.00 & 2.00 & 3.00 & 2.00 \\ 
%500 & 5.12 & 5.92 & 4.09 & 0.87 & 0.97 & 0.53 & 0.00 & 0.00 & 0.00 & 5.00 & 6.00 & 4.00 \\ 
%2000 & 10.60 & 11.70 & 6.81 & 2.67 & 2.83 & 1.05 & 0.10 & 0.11 & 0.00 & 10.00 & 12.00 & 7.00 \\ 
%   \hline
%\end{tabular}
%\caption{Univariate: 12 components - i.e. $\Big(x_i\sim \sum_{j=0}^{11} \frac{1}{12} N(3j,1) \Big)$.}
%\label{tab:univK12}
%\end{table}

\begin{table}[h]
\centering
\caption{Posterior indexes for the multiariate scenario with $k=3$ and $d=4$.}\label{tab:univk3d4}
\begin{tabular}{ccccccc}
%\hline 
 & \multicolumn{2}{c}{Loss-Based} & \multicolumn{2}{c}{Uniform} & \multicolumn{2}{c}{Poi(1)} \\ 
\hline 
$n$ & Mode  & C.I. & Mode  & C.I. & Mode & C.I. \\ 
\hline 
50 & 2.04  & (2.00, 3.54) & 2.13  & (2.00, 4.43) & 2.02  & (2.00, 3.12) \\ 
%\hline 
100 & 2.61   & (2.18, 3.87) & 2.73  & (2.30, 4.44) & 2.58 & (2.16, 3.58) \\ 
%\hline 
500 & 3.00  & (3.00, 3.91) & 3.01 & (3.00, 4.01) & 3.00  & (3.00, 3.15) \\ 
%\hline 
2000 & 3.00  & (3.00, 3.14) & 3.00  & (3.00, 3.33) & 3.00  & (3.00, 3.05) \\ 
%\hline 
\end{tabular} 
\end{table}

%\begin{table}[ht]
%\centering
%\scriptsize
%\begin{tabular}{rrrrrrrrrrrrr}
%  \hline
%n & Mn OBJ & Mn UNI & Mn POI & Sd OBJ & Sd UNI & Sd POI & Fr OBJ & Fr UNI & Fr POI & Md OBJ & Md UNI & Md POI \\ 
%  \hline
%50 & 2.04 & 2.13 & 2.02 & 0.20 & 0.34 & 0.14 & 0.04 & 0.13 & 0.02 & 2.00 & 2.00 & 2.00 \\ 
%100 & 2.61 & 2.73 & 2.58 & 0.49 & 0.47 & 0.50 & 0.61 & 0.71 & 0.58 & 3.00 & 3.00 & 3.00 \\ 
%500 & 3.01 & 3.01 & 3.00 & 0.10 & 0.10 & 0.00 & 0.99 & 0.99 & 1.00 & 3.00 & 3.00 & 3.00 \\ 
%2000 & 3.00 & 3.00 & 3.00 & 0.00 & 0.00 & 0.00 & 1.00 & 1.00 & 1.00 & 3.00 & 3.00 & 3.00 \\ 
%   \hline
%\end{tabular}
%\caption{Multivariate: three components with $d=4$ - i.e. $\Big(x_i\sim \frac{1}{3}N(m,I)+\frac{1}{3}N(0,I)+\frac{1}{3}N(-m,I) \Big)$, with $m=(3/\sqrt{4},\ldots,3/\sqrt{4})$}
%\label{tab:univk3d4}
%\end{table}

\begin{table}[h]
\centering
\caption{Posterior indexes for the multiariate scenario with $k=3$ and $d=8$.}\label{tab:univk3d8}
\begin{tabular}{ccccccc}
%\hline 
 & \multicolumn{2}{c}{Loss-Based} & \multicolumn{2}{c}{Uniform} & \multicolumn{2}{c}{Poi(1)} \\ 
\hline 
$n$ & Mode  & C.I. & Mode  & C.I. & Mode  & C.I. \\ 
\hline 
50 & 2.09  & (2.00, 3.26) & 2.16  & (2.00, 4.13) & 2.06  & (2.00, 3.11) \\ 
%\hline 
100 & 2.56   & (2.22, 3.71) & 2.63 & (2.26, 3.94) & 2.49  & (2.22, 3.43) \\ 
%\hline 
500 & 3.00  & (3.00, 3.26) & 3.00  & (3.00, 4.00) & 3.00  & (3.00, 3.01) \\ 
%\hline 
2000 & 3.00  & (3.00, 3.02) & 3.00  & (3.00, 3.09) & 3.00  & (3.00, 3.01) \\ 
%\hline 
\end{tabular} 
\end{table}

%\begin{table}[ht]
%\centering
%\scriptsize
%\begin{tabular}{rrrrrrrrrrrrr}
%  \hline
%n & Mn OBJ & Mn UNI & Mn POI & Sd OBJ & Sd UNI & Sd POI & Fr OBJ & Fr UNI & Fr POI & Md OBJ & Md UNI & Md POI \\ 
%  \hline
%50 & 2.09 & 2.16 & 2.06 & 0.29 & 0.37 & 0.24 & 0.09 & 0.16 & 0.06 & 2.00 & 2.00 & 2.00 \\ 
%100 & 2.56 & 2.63 & 2.49 & 0.50 & 0.49 & 0.50 & 0.56 & 0.63 & 0.49 & 3.00 & 3.00 & 2.00 \\ 
%500 & 3.00 & 3.00 & 3.00 & 0.00 & 0.00 & 0.00 & 1.00 & 1.00 & 1.00 & 3.00 & 3.00 & 3.00 \\ 
%2000 & 3.00 & 3.00 & 3.00 & 0.00 & 0.00 & 0.00 & 1.00 & 1.00 & 1.00 & 3.00 & 3.00 & 3.00 \\ 
%   \hline
%\end{tabular}
%\caption{Multivariate: three components with $d=8$ - i.e. $\Big(x_i\sim \frac{1}{3}N(m,I)+\frac{1}{3}N(0,I)+\frac{1}{3}N(-m,I) \Big)$, with $m=(3/\sqrt{8},\ldots,3/\sqrt{8})$}
%\label{tab:univk3d8}
%\end{table}

\begin{table}[h]
\centering
\caption{Posterior indexes for the multiariate scenario with $k=3$ and $d=12$.}\label{tab:univk3d12}
\begin{tabular}{ccccccc}
%\hline 
 & \multicolumn{2}{c}{Loss-Based} & \multicolumn{2}{c}{Uniform} & \multicolumn{2}{c}{Poi(1)} \\ 
\hline 
$n$ & Mode  & C.I. & Mode  & C.I. & Mode  & C.I. \\ 
\hline 
50 & 2.05  & (2.00, 3.13) & 2.08  & (2.00, 3.61) & 2.02  & (2.00, 3.06) \\ 
%\hline 
100 & 2.23  & (2.11, 3.33) & 2.30  & (2.11, 3.44) & 2.22  & (2.11, 3.00) \\ 
%\hline 
500 & 3.00  & (3.00, 3.08) & 3.00  & (3.00, 3.98) & 3.00  & (3.00, 3.01) \\ 
%\hline 
2000 & 3.00  & (3.00, 3.00) & 3.00  & (3.00, 3.00) & 3.00  & (3.00, 3.00) \\ 
%\hline 
\end{tabular} 
\end{table}

%\begin{table}[ht]
%\centering
%\scriptsize
%\begin{tabular}{rrrrrrrrrrrrr}
%  \hline
%n & Mn OBJ & Mn UNI & Mn POI & Sd OBJ & Sd UNI & Sd POI & Fr OBJ & Fr UNI & Fr POI & Md OBJ & Md UNI & Md POI \\ 
%  \hline
%50 & 2.05 & 2.08 & 2.02 & 0.22 & 0.27 & 0.14 & 0.00 & 0.00 & 0.00 & 2.00 & 2.00 & 2.00 \\ 
%100 & 2.23 & 2.30 & 2.22 & 0.42 & 0.46 & 0.42 & 0.00 & 0.00 & 0.00 & 2.00 & 2.00 & 2.00 \\ 
%500 & 3.00 & 3.00 & 3.00 & 0.00 & 0.00 & 0.00 & 0.00 & 0.00 & 0.00 & 3.00 & 3.00 & 3.00 \\ 
%2000 & 3.00 & 3.00 & 3.00 & 0.00 & 0.00 & 0.00 & 0.00 & 0.00 & 0.00 & 3.00 & 3.00 & 3.00 \\ 
%   \hline
%\end{tabular}
%\caption{Multivariate: three components with $d=12$ - i.e. $\Big(x_i\sim \frac{1}{3}N(m,I)+\frac{1}{3}N(0,I)+\frac{1}{3}N(-m,I) \Big)$, with $m=(3/\sqrt{12},\ldots,3/\sqrt{12})$}
%\label{tab:univk3d12}
%\end{table}

From the simulation study, in particular when comparing the three priors, we can observe the following. In the univariate cases, the uniform prior tends to have the largest size of the posterior credible intervals. This is expected as the mass on each value of $k$ is constant. Whilst this may be regarded as a disadvantage when $k$ is relatively low, we see that, besides the case of $k=12$ and $n=50$ or $n=100$, the posterior credible intervals always contain the true value of the number of components. At the opposite, we see that the Poisson prior (with parameter equal to 1), tends to return the smallest credible intervals. However, due to its fast decrease to zero for increasing $k$, there are already cases for $k=4$ where, if the information in the sample is not sufficient, it fails to contain the true parameter value. The proposed loss-based prior, on the other hand, has a behaviour that is somewhere between the others. In fact, the credible interval size is closer to the Poisson prior (rather than the uniform prior) and, with the exception of $k=12$ and $n=50,100,500$, it always contain the true value.

When we compare the priors in the multivariate setting, we note very few differences. Posterior average modes and credible intervals have similar values, with negligible impact for difference dimensionality.

\end{document}